# NEFFY 2.0: A Breathing Companion Robot: User-Centered Design and Findings from a Study with Ukrainian Refugees

Ilona Buchem
Berlin University of Applied Sciences
Berlin Germany
buchem@bht-berlin.de

Jessica Kazubski
Berlin University of Applied Sciences
Berlin Germany
jessica_kazubski@web.de

Charly Goerke
Berlin University of Applied Sciences
Berlin Germany
goerke.charly@gmail.com

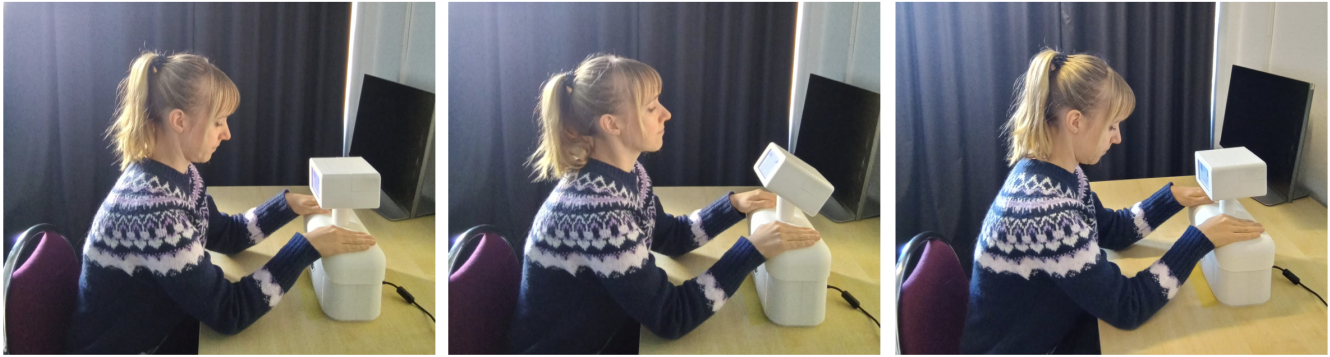

Figure 1: NEFFY 2.0 – A Slow-Paced Breathing Companion Robot.

## Abstract

This paper presents the design of NEFFY 2.0, a social robot designed as a haptic slow-paced breathing companion for stress reduction, and reports findings from a mixed-methods user study with 14 refugees from Ukraine. Developed through a user-centered design process, NEFFY 2.0 builds on NEFFY 1.0 and integrates embodiment and multi-sensory interaction to provide low-threshold, accessible guidance of slow-paced breathing for stress relief, which may be particularly valuable for individuals experiencing prolonged periods of anxiety. To evaluate effectiveness, an experimental comparison of a robot-assisted breathing intervention versus an audio-only condition was conducted. Measures included subjective ratings and physiological indicators, such as heart rate (HR), heart rate variability (HRV) using RMSSD parameter, respiratory rate (RR), and galvanic skin response (GSR), alongside qualitative data from interviews exploring user experience and perceived support. Qualitative findings showed that NEFFY 2.0 was perceived as intuitive, calming and supportive. Survey results showed a substantially larger effect in significant reduction of perceived stress in the NEFFY 2.0 condition compared to audio-only. Physiological data reveled mixed results combined with large inter-personal variability. Three patterns of breathing practice with NEFFY 2.0 were identified using k-means clustering. Despite the small sample size, this study makes a novel contribution by providing empirical evidence of stress reduction in a vulnerable population through a direct comparison of robot-assisted and non-robot conditions. The findings position NEFFY 2.0 as a promising low-threshold tool that supports stress relief and contributes to the vision of HRI empowering society.

## CCS Concepts

• **General and reference** → **Design**; • **Computer systems organization** → **Robotics**; • **Human-centered computing** → **User studies**.

## Keywords

human-robot interaction (HRI), social robots (SR), stress reduction, slow breathing, physiological stress indicators, NEFFY robot

## 1 Introduction and Related Work

Research on socially assistive robots (SARs) highlights their potential to support human wellbeing, mental health, social interaction and stress regulation, decrease anxiety and loneliness, improve mood, comfort and quality of life, reduce pain across the lifespan [1–4]. While robotic interventions for wellbeing have typically relied on visual or auditory guidance [5], tactile and haptic modalities remain still underexplored, despite evidence that physical cues, such as rhythmic motion mimicking breathing, can transfer arousal through touch [6], decrease arousal [7], reduce anxiety [8] and increase calmness and contentment [9]. Even fewer studies with SARs included vulnerable populations, such as refugees, who may especially benefit from supportive physical cues. Prolonged exposure to displacement can place a substantial psychological burden on persons who have fled conflict, including post-traumatic stress disorder (PTSD), depression and anxiety [10]. Preliminary studies show that SARs can benefit vulnerable populations, including adult immigrants, through reduction of depressive symptoms, improvement in emotional well-being, and decrease in loneliness [11]. Slow-paced breathing has been considered an effective, low-threshold technique for reducing stress and restoring balance [12–16]. SARs can be designed and applied to guide breathing through an embodied, multisensory and paced interaction with integrated haptic, visual and auditory modalities to decrease the effects of stress [8, 9].

NEFFY 2.0 was developed in the Communications Lab at Berlin University of Applied Sciences, Germany as a haptic breathing companion guiding slow-paced breathing through soft, rhythmic motion and multisensory feedback. NEFFY 2.0 builds on the first prototype NEFFY 1.0 and exploratory studies which showed positive effects of NEFFY-guided breathing practice on perceived calmness and reductions in self-reported anxiety, alongside high acceptance and perceived co-presence [9]. NEFFY 2.0 was designed as a social companion [4] with an approachable form, gentle, predictable motion, low linguistic demand, and a user-friendly interaction.

Prior studies with NEFFY did not include a control group, nor physiological stress measurement, and the sample consisted primarily of students [9]. This paper reports findings from a mixed-methods, within-subjects user study with a vulnerable group of adult refugees from Ukraine. Participants completed a NEFFY-assisted and an audio-only breathing session, accompanied by subjective and physiological stress measurement. The study investigates the effects of a robot-assisted compared to audio-only experience.

## 2 User Centered Design

The development of NEFFY 2.0 followed user-centered design (UCD) approach to ensure that the robot's form, functionality, and interaction patterns aligned with users' needs [17–19]. NEFFY 2.0 is part of ongoing research, which investigates how social robots can support stress relief and wellbeing. Insights from studies with NEFFY 1.0 indicated that users experience the interaction as calming and pleasant, highlighting the potential of a companion robot to facilitate slow-paced breathing practice for stress relief [9].

The work was carried out by a multidisciplinary team with a hardware engineer (third author), a software engineer (second author), and an HRI researcher (first author), and structured into three interdependent activities: technical construction (design of 3D-printed components, integration of hardware modules, realization of the robot's motion); interaction design (display, facial animations, audio output, coordinated motor control); and user research in continuous exchange, enabling iterative refinement across mechanical and interaction layers. In prototyping phases, alternative technical and interaction designs were developed, tested, and reviewed with potential users. Feedback guided decisions toward solutions that best matched user expectations regarding comfort, intuitiveness, and perceived support.

### 2.1 Design Goals

Design of NEFFY 2.0 was shaped by insights from prior studies with NEFFY 1.0 [9] and requirements for a low-threshold breathing companion robot [8, 20]. A central objective was to preserve the recognizable form, characteristic movement patterns, and interaction logic of NEFFY 1.0, while upgrading the system in ways that would enhance stability, usability, and expressiveness. Maintaining continuity allowed to build on interaction elements that users already perceived as calming and intuitive, while systematically refining aspects that limited the application of the first prototype.

The second design goal was the improvement of mechanical stability. User studies with NEFFY 1.0 showed that the torso and head movement system needed to tolerate sustained physical contact and repeated actuation. The goal for NEFFY 2.0 was to achieve greater mechanical robustness, smoother and quieter motion, and to ensure that haptic guidance remained consistent even when users put some pressure resting their hands on the robot.

The third goal was to expand NEFFY's interaction capabilities beyond purely visual and haptic cues based on user feedback. NEFFY 2.0 was designed to communicate verbally, provide spoken prompts, and guide users through the exercise in an accessible manner and in multiple languages. Also, expressive facial animations were introduced to make the interaction more engaging and intuitive.

Finally, NEFFY 2.0 was developed to enable independent, operator-free use. Whereas NEFFY 1.0 required a human facilitator to explain the exercise and to set up the session, NEFFY 2.0 can function autonomously, allowing users to switch on/off and use the touchscreen to select language, duration of the exercise and adjust volume.

### 2.2 Improvements

NEFFY 2.0 was modeled in SOLIDWORKS 2023 and fabricated using PLA on a Prusa MK3S 3D printer. Several changes were introduced compared to NEFFY 1.0. First, the width and height were slightly increased to provide a larger and more comfortable hand contact area. Second, NEFFY 2.0 replaced a 3D-printed crank–slider mechanism with a belt-driven system powered by a Dynamixel XL430-W250 servo motor to generate smoother shoulder motion, allowing for vertical translation of up to 25 mm. In cases of excessive pressure, the motor automatically shuts down and reinitializes safely once the force is removed. Third, the head motion was redesigned. The pin-on-servo-horn mechanism of NEFFY 1.0 was replaced with a pair of 3D-printed spur gears to generate smooth nodding of up to 40 degrees. The head shell now consists of three lightweight magnetic sections for an easy assembly and maintenance.

Finally, structural changes were implemented to improve access to electronic components, facilitating software updates and maintenance. The front torso houses the speaker and shoulder drive, while the rear torso contains the Raspberry Pi 4B, motor controllers, voltage regulation, and power interface. Cable routing was simplified, and the system now includes an external on/off switch, power jack, and modular shoulder plates with magnets. The dual-screen eye configuration from NEFFY 1.0 was replaced by a single integrated touchscreen that displays the robot's full face and supports user input. The audio system was upgraded from an internal USB speaker to a front-mounted external speaker for clearer sound.

## 2.3 User Interaction

NEFFY 2.0 introduces a reworked interaction design that integrates improvements on the display and motor control layers. A virtual Python environment was created on the Raspberry Pi, GPIO pins were configured for external components, and scripts were set to launch automatically upon boot. Due to the ILI9488 display driver, dedicated libraries, e.g. Adafruit RGB Display, were adapted and extended to enable rendering graphics and processing touch input. The display now presents dynamic facial expressions and handles user input during configuration steps. Audio output was implemented using the Pygame library, allowing playback of breathing sounds, background music, and spoken prompts. Volume control is accessible directly via touchscreen, and the audio output is routed through the AUX port of the Raspberry Pi to the upgraded external speaker. Dynamixel motors controlling the head and shoulder movements were programmed using the Dynamixel SDK and supplemental modules. All modules were integrated into a central script that structures the interaction into logical phases, including greeting, configuration, breathing guidance, and farewell. This modular architecture ensures reliable synchronization across components and allows rapid updates to specific behaviors. NEFFY 2.0 guides users through a sequence of five phases: Startup, Greeting, Settings, Breathing, Ending. By harmonizing haptic motion, expressive visuals, and auditory cues, NEFFY 2.0 now offers an intuitive breathing experience to benefit users through simple and calming interaction.

## 3 User Study

This study explore the effects of a slow-paced breathing exercise, with and without a robot, on subjective and physiological stress indicators in a vulnerable population of Ukrainian refugees. The study was reviewed and approved by the Ethics Board of the Berlin University of Applied Sciences (BHT).

### 3.1 Study Design

The study followed a mixed-methods, within-subjects experimental design in which all participants completed two five-minute breathing sessions: one guided by the NEFFY 2.0 robot and one guided by the identical audio track alone with a 5-minute break between conditions to facilitate a washout effect. This approach allowed each participant to serve as their own control [21]. The order was counterbalanced to minimize order effects [22]. However, residual carryover or novelty effects of the robot cannot be fully excluded.

### 3.2 Participants

Recruitment was conducted through a local network involving a non-profit social welfare organization. The inclusion criteria were being an adult and a Ukrainian refugee. Participation was voluntary, and no personally identifying information was collected (each participant was assigned an ID). We recruited 20 adult Ukrainian refugees. However, due to some no-shows and incomplete data, the final sample included 14 refugees, 10 women and 4 men (age range 22–85, average age 52). This sample size aligns with exploratory HRI aiming to generate first insights in vulnerable communities. We acknowledge that stress levels vary among refugees and that refugee status alone does not imply elevated baseline stress. The observed inter-individual variability in our sample reflects this heterogeneity.

The majority of the participants (9/14) had no prior experience with breathing practice and none had experience with social robots. Russian was the preferred language for 11/14, Ukrainian for 3/14 participants. All study materials, including consent forms, robot speech, surveys, displays and interview guides, were provided in Russian and Ukrainian to ensure linguistic accessibility [23].

### 3.3 Procedure

Each participant completed two sessions, i.e. one with NEFFY 2.0 and one with an audio track, both on the same day. In the robotic condition, NEFFY 2.0 provided synchronized haptic (shoulder actuation), visual (animated facial cues), and auditory (robot's voice and chime sounds) guidance. In the audio condition, the same instructions were delivered via audio only. Each session lasted approx. 30 minutes with 5 minutes breathing practice. Each session followed a standardized procedure: Physiological data and self-reported stress were collected immediately before and after each exercise. A short semi-structured interview followed each session to capture subjective experience. Upon arrival, participants were welcomed and informed about the study. Physiological sensors were attached, i.e. Shimmer3 GSR+ sensor for heart activity (PPG) and skin conductance (GSR) [24], and Vernier Go Direct Respiration Belt for respiratory rate [25]. The Shimmer3 PPG clip was placed on the earlobe, the GSR electrodes on the middle and ring finger of the non-dominant hand. The respiration belt was positioned around the lower ribcage above the diaphragm. After sensor calibration, synchronized time logging began. Participants completed the baseline survey (STAI 6 and demographics), and the post-survey (STAI-6, NASA-TLX, and MDMT in the robot condition) after the exercise, followed by a short interview. At the end, sensors were removed.

### 3.4 Measures

Using sensors in each session, four established physiological stress indicators were collected: Heart Rate (HR) as a number of beats per minute [26], Heart Rate Variability (HRV) as a variation in inter-beat intervals using Root Mean Square of Successive Differences (RMSSD) as parameter [26, 27], Respiratory Rate (RR) as breaths per minute [28], and Galvanic Skin Response (GSR) as skin conductance influenced by sweat gland activity [29]. Increases in HR, RR, and GSR and decreases in HRV are commonly interpreted as elevated physiological arousal associated with stress, and these indicators allow for a multidimensional measurement of stress [30].

Self-reported measures included three validated scales, i.e. 1. State–Trait Anxiety Inventory (STAI-6) [31] (state anxiety; pre- and post-exercise) with Cronbach's $\alpha$: audio pre = .81, audio post = .62, NEFFY pre = .77, NEFFY post = .65; 2. Multidimensional Measure of Trust (MDMT) [32] (trust in the robot; robot condition only) with Cronbach's $\alpha$ = 0.76; and 3. NASA Task Load Index (NASA-TLX) [33] (perceived workload; post-exercise) with Cronbach's $\alpha$: Neffy = .58, audio = .75. All instruments were available in Russian and Ukrainian to ensure linguistic and cultural accessibility.

## 4 Results

Data analysis was conducted in a multi-stage procedure using quantitative and qualitative methods to explore the effects of breathing with NEFFY 2.0 in comparison to the non-robotic intervention.

Table 1: Average values and change from minute 1 to 5 (Δ).

| Parameter | Neffy 2.0 | Audio | Δ Neffy | Δ Audio |
|---|---|---|---|---|
| HR (BPM) | 73.18 | 70.99 | 3.22 | 2.16 |
| HRV (ms) | 75.36 | 91.91 | -10.64 | +3.42 |
| GSR ($\mu S$) | 0.856 | 1.048 | -0.14 | -0.21 |

### 4.1 Subjective Stress

Analysis of STAI-6 scores revealed statistically significant reductions in perceived stress in both conditions CR (robot) and CA (audio), with substantially stronger effects in CR. In CA, participants showed a modest decrease from pre- to post-exercise ($M_{pre}$ = 39.25, $M_{post}$ = 34.73), $t(13)$ = 2.17, $p$ = .049, Cohen's $d$ = −0.58. In contrast, CR yielded a markedly larger reduction ($M_{pre}$ = 41.62, $M_{post}$ = 30.92), $t(13)$ = 4.96, $p$ = .0003, Cohen's $d$ = −1.33, representing a large effect. The comparison of change scores in CR and CA showed a trend toward greater improvement with CR ($M_{\Delta Audio}$ = −4.52 vs. $M_{\Delta NEFFY}$ = −10.70). However, this difference was not statistically significant, $t(13)$ = 2.02, $p$ = .064.

### 4.2 Physiological Indicators

Physiological data, i.e. heart rate (HR), heart rate variability (HRV), respiratory rate (RR), galvanic skin response (GSR), were processed following standard quality control procedures: PPG signals were visually inspected for motion artifacts, noisy segments were excluded, and HRV was computed using the RMSSD metric over the 5-minute breathing interval. GSR signals were checked for signal loss and analyzed as means across the same interval. Missing or invalid data was excluded from the analyses. The results were heterogeneous with no significant differences between conditions (ANOVA, paired t-tests). HR was higher in CR and the average difference between minutes 1 and 5 was Δ = 3.22 BPM (in CA Δ = 2.16 BPM). HRV was higher in CA (91.91 ms) with Δ = 3.42 ms. In CR, HRV was lower (75.36 ms), with Δ = -10.64 ms. GSR was lower in CR (0.856 $\mu S$) and decreased by -0.14 $\mu S$. In CA, GSR was higher with 1.048 $\mu S$ and Δ = -0.21 $\mu S$ (Table 1). In both conditions, RR approached the target rate of six breaths per minute. The average RR was 6.20 in CR, and 6.50 in CA. The average deviation from the target value was lower in CR (0.57 BPM) compared to CA (1.05 BPM).

### 4.3 Cluster Analysis

To explore robot-practice patterns, we conducted an exploratory cluster analysis combining subjective (STAI-6, NASA-TLX, MDMT) and physiological (HR, HRV, GSR, RR) indicators. All variables were standardized. A three-cluster solution was identified using k-means clustering, based on silhouette coefficient analysis combined with visual inspection [34]. Cluster 1 indicates subjective and objective stress relief (STAI-6: −6.95, GSR: −0.30), despite the highest workload (NASA-TLX: 8.26), slight increase in HR (0.78) and RR (0.30), and decrease in HRV (−5.26). Cluster 2 shows the highest subjective stress relief (−7.41) combined with the lowest workload (3.52), decrease in GSR (−0.07) and RR (0.21), increase in HR (6.90) and the largest drop in HRV (−13.86). Cluster 3 shows the highest trust (MDMT: 7.00) combined with high subjective and objective stress reduction (STAI-6: −6.66; HR: −0.52), low workload (NASA-TLX: 4.33), but an increase in RR (3.32) and GSR (0.24), and a drop in HRV (−11.18). Given the small samples size, the k-means clustering results are exploratory and serve to inform future research.

### 4.4 Qualitative Findings

Data from the interviews was transcribed, translated and analyzed using structuring qualitative content analysis [35]. All 14 interviews revealed positive experiences with the NEFFY 2.0. 11 participants explicitly reported feeling more relaxed after the exercise, describing the robot as calming, intuitive, or supportive. Only 3 participants mentioned no noticeable effect, linking it to own illness or fatigue (e.g. shortness of breath due to a cold, having walked just before the session). One participant described NEFFY's breathing as unfamiliar. Nobody reported confusion about inhales or exhales in the robot condition. In contrast, in the audio-only condition, 7 participants reported difficulties related to uncertainty about the breathing rhythm ("I didn't know when to inhale," "I could not catch the rhythm"). Comments from the audio-condition expressed a need for clearer cues. While in both conditions, participants expressed appreciation toward the interventions, NEFFY 2.0 was perceived as more intuitive compared to the audio-only guidance.

## 5 Discussion

This study provides evidence that multisensory robot-guided breathing practice contributes to higher perceived stress reduction in comparison to audio-only guidance. Qualitative findings show that the combination of haptic guidance, visual and audio cues supported stress reduction. Also, the large effect size observed for NEFFY 2.0 in STAI-6 ($d$ = −1.33) which exceeded the audio condition, may be attributed to the multisensory guidance of the robot. While previous work in SARs for slow-breathing demonstrated positive effects [8, 9], these studies did not include direct comparisons with other types of interventions (e.g. audio) nor examined the effects for vulnerable groups (e.g. refugees). This makes the present findings a novel contribution to prior research. However, despite counterbalancing, repeated same-day self-reports may have reduced stress in the second condition, potentially affecting condition comparisons. Significant subjective stress reduction in both conditions aligns with evidence on the effectiveness of slow breathing [12, 13, 15, 16]. Physiological indicators showed only few clearly positive effects (e.g. reaching the RR target of six breaths per minute) despite positive self-reported results. This aligns with prior findings that subjective and objective measures do not always correspond with each other [26, 29, 30]. The cluster analysis revealed three distinct response patterns. The results indicate the need of multidimensional stress assessment [30, 36]. While the small sample size limits statistical generalization, methodologically this work extends prior research [8, 9] and contributes to HRI empowering society.

Future research with larger samples of vulnerable populations can build on this work by investigating adaptive interventions of robot-supported breathing practice and examining how haptic interaction may contribute to stress regulation over time.

## 6 Acknowledgments

We thank Pflege & Wohnen Sunpark – Johannesstift Diakonie in Berlin, Germany for the support in recruiting study participants.